\documentclass[prl,twocolumn,aps,amsmath,amssymb,nofootinbib,preprintnumbers]
{revtex4}

\usepackage{graphicx}
\usepackage{dcolumn}
\usepackage{bm}
\usepackage{amsmath}
\usepackage{amsfonts}

\def\ls{\mathrel{\lower4pt\vbox{\lineskip=0pt\baselineskip=0pt
           \hbox{$<$}\hbox{$\sim$}}}}
\def\gs{\mathrel{\lower4pt\vbox{\lineskip=0pt\baselineskip=0pt
           \hbox{$>$}\hbox{$\sim$}}}}
\def\drawbox#1#2{\hrule height#2pt

\hbox{\vrule width#2pt height#1pt \kern#1pt
              \vrule width#2pt}
              \hrule height#2pt}

\def\Asym#1#2{\vcenter{\vbox{\drawbox{#1}{#2}
              \kern-#2pt       
              \drawbox{#1}{#2}}}}

\begin{document}


\title{Landau levels and Riemann zeros
}

\author{Germ\'an Sierra$^1$  and Paul K. Townsend $^2$}

\affiliation{$1$ Instituto de F\'{\i}sica Te\'orica, CSIC-UAM, Madrid, Spain \\
$2$ Department of Applied
Mathematics and Theoretical Physics \\
Centre for Mathematical
Sciences, University of Cambridge, UK}


\bigskip\bigskip\bigskip\bigskip

%
\font\numbers=cmss12
\font\upright=cmu10 scaled\magstep1
\def\stroke{\vrule height8pt width0.4pt depth-0.1pt}
\def\topfleck{\vrule height8pt width0.5pt depth-5.9pt}
\def\botfleck{\vrule height2pt width0.5pt depth0.1pt}
\def\Zmath{\vcenter{\hbox{\numbers\rlap{\rlap{Z}\kern
0.8pt\topfleck}\kern 2.2pt
                   \rlap Z\kern 6pt\botfleck\kern 1pt}}}
\def\Qmath{\vcenter{\hbox{\upright\rlap{\rlap{Q}\kern
                   3.8pt\stroke}\phantom{Q}}}}
\def\Nmath{\vcenter{\hbox{\upright\rlap{I}\kern 1.7pt N}}}
\def\Cmath{\vcenter{\hbox{\upright\rlap{\rlap{C}\kern
                   3.8pt\stroke}\phantom{C}}}}
\def\Rmath{\vcenter{\hbox{\upright\rlap{I}\kern 1.7pt R}}}
\def\Z{\ifmmode\Zmath\else$\Zmath$\fi}
\def\Q{{\bf Q}}
\def\P{{\bf P}}
\def\L{{\bf L}}
\def\M{{\bf M}}
\def\N{{\bf N}}
\def\C{\ifmmode\Cmath\else$\Cmath$\fi}
\def\R{\ifmmode\Rmath\else$\Rmath$\fi}
\def\H{{\cal H}}
\def\NN{{\cal N}}
\def\cl{{\rm cl}}
\def\RS{{\rm RS}}
\def\E{{\rm E}}

\begin{abstract}

The number $N(E)$ of complex zeros of the Riemann zeta function  with positive imaginary part 
less than $E$ is the sum of a  `smooth'  function $\bar N(E)$ and a `fluctuation'. Berry and Keating have shown that the asymptotic expansion of $\bar N(E)$ counts states of positive energy less than $E$ in a  `regularized'  semi-classical model with classical Hamiltonian $H=xp$.  For a different regularization, Connes has shown that it counts states `missing'  from a continuum.  Here we show how  the `absorption  spectrum'  model  of  Connes  emerges as the lowest  Landau level  limit of a specific  quantum mechanical model for a charged particle on a planar  surface  in an electric  potential and uniform  magnetic field.  We suggest a role for the higher Landau levels in the fluctuation part of  $N(E)$. 
 
 \end{abstract}

\preprint{IFT-UAM/CSIC08-26, DAMTP-2008}

\maketitle

\vskip 0.2cm

%
%
%
%
\def\oti{{\otimes}}
\def\lb{ \left[ }
\def\rb{ \right]  }
\def\tilde{\widetilde}
\def\bar{\overline}
\def\hat{\widehat}
\def\*{\star}
\def\[{\left[}
\def\]{\right]}

%
%
\def\zb{{\bar{z} }}
\def\z{{\bf z}}
\def\x{{\bf x}}

\def\zbar{{\bar{z} }}
\def\d{\partial}
\def\vev#1{\langle #1 \rangle}
\def\ket#1{ | #1 \rangle}
\def\rvac{\hbox{$\vert 0\rangle$}}
\def\lvac{\hbox{$\langle 0 \vert $}}
\def\2pi{\hbox{$2\pi i$}}
\def\e#1{{\rm e}^{^{\textstyle #1}}}
\def\grad#1{\,\nabla\!_{{#1}}\,}
\def\dsl{\raise.15ex\hbox{/}\kern-.57em\partial}
\def\Dsl{\,\raise.15ex\hbox{/}\mkern-.13.5mu D}
%
%
\def\ga{\gamma}     \def\Ga{\Gamma}
\def\be{\beta}
\def\al{\alpha}
\def\ep{\epsilon}
\def\vep{\varepsilon}
\def\dep{d}
\def\arc{{\rm Arctan}}
\def\la{\lambda}    \def\La{\Lambda}
\def\de{\delta}     \def\De{\Delta}
\def\om{\omega}     \def\Om{\Omega}
\def\sig{\sigma}    \def\Sig{\Sigma}
\def\vphi{\varphi}
\def\sign{{\rm sign}}
\def\he{\hat{e}}
\def\hf{\hat{f}}
\def\hg{\hat{g}}
\def\ha{\hat{a}}
\def\hb{\hat{b}}
\def\hx{\hat{x}}
\def\hp{\hat{p}}
\def\f{{\bf f}}
\def\g{{\bf g}}
\def\a{{\bf a}}
\def\b{{\bf b}}
\def\fl{{\rm fl}}
\def\sm{{\rm sm}}
\def\QM{{\rm QM}}
\def\tV{{\tilde{V}}}
\def\tH{{\tilde{H}}}

%
%
\def\CA{{\cal A}}   \def\CB{{\cal B}}   \def\CC{{\cal C}}
\def\CD{{\cal D}}   \def\CE{{\cal E}}   \def\CF{{\cal F}}
\def\CG{{\cal G}}   \def\CH{{\cal H}}   \def\CI{{\cal J}}
\def\CJ{{\cal J}}   \def\CK{{\cal K}}   \def\CL{{\cal L}}
\def\CM{{\cal M}}   \def\CN{{\cal N}}   \def\CO{{\cal O}}
\def\CP{{\cal P}}   \def\CQ{{\cal Q}}   \def\CR{{\cal R}}
\def\CS{{\cal S}}   \def\CT{{\cal T}}   \def\CU{{\cal U}}
\def\CV{{\cal V}}   \def\CW{{\cal W}}   \def\CX{{\cal X}}
\def\CY{{\cal Y}}   \def\CZ{{\cal Z}}

\def\Hp{{\mathbb{H}^2_+}} 
\def\Hm{{\mathbb{H}^2_-}}
\def\tphi{{\tilde \phi}}

\def\rvac{\hbox{$\vert 0\rangle$}}
\def\lvac{\hbox{$\langle 0 \vert $}}
\def\comm#1#2{ \BBL\ #1\ ,\ #2 \BBR }
\def\2pi{\hbox{$2\pi i$}}
\def\e#1{{\rm e}^{^{\textstyle #1}}}
\def\grad#1{\,\nabla\!_{{#1}}\,}
\def\dsl{\raise.15ex\hbox{/}\kern-.57em\partial}
\def\Dsl{\,\raise.15ex\hbox{/}\mkern-.13.5mu D}
%
%
%
\font\numbers=cmss12
\font\upright=cmu10 scaled\magstep1
\def\stroke{\vrule height8pt width0.4pt depth-0.1pt}
\def\topfleck{\vrule height8pt width0.5pt depth-5.9pt}
\def\botfleck{\vrule height2pt width0.5pt depth0.1pt}
\def\Zmath{\vcenter{\hbox{\numbers\rlap{\rlap{Z}\kern
0.8pt\topfleck}\kern 2.2pt
                   \rlap Z\kern 6pt\botfleck\kern 1pt}}}
\def\Qmath{\vcenter{\hbox{\upright\rlap{\rlap{Q}\kern
                   3.8pt\stroke}\phantom{Q}}}}
\def\Nmath{\vcenter{\hbox{\upright\rlap{I}\kern 1.7pt N}}}
\def\Cmath{\vcenter{\hbox{\upright\rlap{\rlap{C}\kern
                   3.8pt\stroke}\phantom{C}}}}
\def\Rmath{\vcenter{\hbox{\upright\rlap{I}\kern 1.7pt R}}}
\def\Z{\ifmmode\Zmath\else$\Zmath$\fi}
\def\C{\ifmmode\Cmath\else$\Cmath$\fi}
\def\R{\ifmmode\Rmath\else$\Rmath$\fi}

\def\barray{\begin{eqnarray}}
\def\earray{\end{eqnarray}}
\def\beq{\begin{equation}}
\def\eeq{\end{equation}}

\def\no{\noindent}

\def\s{\sigma}
\def\Ga{\Gamma}

\def\g{{\bf g}}
\def\K{{\cal K}}
\def\I{{\cal I}}

\def\F{{\cal F}}

\def\Im{{\rm Im}}
\def\Re{{\rm Re}}
\def\ti{{\tilde{\phi}}}
\def\tR{{\tilde{R}}}
\def\tS{{\tilde{S}}}
\def\tF{{\tilde{\cal F}}}
\def\om{{\omega_0}} 
\def\oc{{\omega_c}} 
\def\oh{{\omega_h}}

At the begining of the 20th century Polya and Hilbert  
conjectured that the imaginary part 
of the complex zeros of the Riemann zeta function $\zeta(s)$ 
are  the eigenvalues of a self-adjoint operator $H$.  
The existence of such an operator implies the celebrated Riemann hypothesis  
that all complex zeros lie on the `critical line' ${\mathcal Re}(s)=1/2$.  
Support  for this `spectral'  approach did not emerge   until the 1950s,  
when Selberg found a `trace' formula relating the eigenvalues of the Laplacian 
on a compact hyperbolic surface to its geodesics;  there is a strong 
resemblance 
of this formula to the Riemann-Weil  `explicit'  formula relating the Riemann 
zeros to the prime numbers.  Further support came in the 1970s and 80s 
from the 
works of  Montgomery and Odlyzko: assuming  the Riemann hypothesis, 
they showed  
that the local statistics of the Riemann zeros is described  by the Gaussian  
Unitary Ensemble (GUE) of Random Matrix Theory (see e.g.  \cite{Frontiers}).  
Inspired by this, and by analogies between trace formulas of  number 
theory and  the Gutzwiller formula for chaotic dynamical systems, 
Berry conjectured the  existence of a classical chaotic Hamiltonian system with isolated 
periodic orbits, with periods related to the prime numbers, such that the spectrum 
of the quantum theory gives  the complex Riemann zeros \cite{B-chaos}. The GUE statistics
of the zeros requires the Hamiltonian to break time-reversal invariance. 

Berry's conjecture received support in 1999 from the work of Connes  
\cite{Connes}, and Berry and Keating \cite{BK1}, on a semi-classical 
model  for a particle in one dimension with classical Hamiltonian 
$H=xp$. This Hamiltonian breaks time-reversal invariance since 
 $(x,p)\to (x,-p) \Rightarrow H\to -H$.  Classical orbits in this model are unbounded 
 hyperpolas in phase space but they may be converted into closed  orbits by an identification at 
the boundaries of a restricted region of the phase plane that defines 
some `regularization'  of the model. This regularization also makes 
finite the number  of states with energy less than $E$, and this number is related to the 
number  $N(E)$ of complex zeros  of the Riemann zeta function with positive imaginary 
part less than $E$.  The Riemann-van Mangoldt  `counting' formula  states that
\beq
N(E)= \bar N(E) + \frac{1}{\pi} {\mathcal Im}\, 
\log \zeta\left(\frac{1}{2} + iE\right)\, 
\label{ne}
\eeq
where $\bar N(E)$ is a `smoothed'  counting function that gives the average
 number of zeros, which is corrected by the  remaining  `fluctuation' term. 
The smooth term can be written as
\beq\label{barNTheta}
\bar N(E) = \frac{\theta(E)}{\pi} + 1\, , 
\eeq
where (see e.g. \cite{Edwards})
\beq\label{theta}
\theta(E) = {\mathcal Im}\, \log \Gamma\left(\frac{1}{4}+ i\frac{E}{2}\right) 
-  \frac{E}{2} \log\pi \, , 
\eeq
which  is the phase of the zeta function on the critical line. The smooth term
has the following asymptotic expansion for  large $E$:
\beq\label{asymp}
\bar N(E) \sim \frac{E}{2 \pi} \log \frac{E}{2 \pi}  
- \frac{E}{2 \pi} + {\cal O}(1)\, , 
\eeq
while the fluctuation term is of order $\log E$.   In the Berry-Keating model this asymptotic expression  
for $\bar N(E)$  is  recovered from a semi-classical count of states with positive energy less  than $E$;  the correct  ${\cal O}(1)$  term is obtained by taking into account the Maslov phase implied by the identifications that close the hyperbolic orbits \cite{BK1}.  In the Connes model, which has a different `regularization', the same asymptotic expression is found  as a semi-classical count of states missing from a  continuous spectrum. Thus, the Berry-Keating interpretation of $\bar N(E)$ is as an `emission'  spectrum  while the Connes interpretation is as an `absorption' spectrum. 

The first aim of this letter is to show how these semi-classical results can be  recovered by viewing the $H=xp$ model as a lowest Landau level  (LLL) limit of a  quantum mechanical model for a charged  particle on the  $xy$-plane in a constant  uniform perpendicular  magnetic field  of strength $B$, and 
an electric potential $\varphi=- \lambda xy$, where $\lambda$ is a constant.  Let $\mu$ be the particle's mass and $-e$ its charge. In  the Landau gauge  $A=Bxdy$ for the vector potential 1-form $A$, the particle's Lagrangian is
\beq
\L = \frac{\mu}{2} (\dot{x}^2 + \dot{y}^2)
-  \; \frac{ e B}{c} \;   \dot{y} x  -e \, \lambda \,   x y \, . 
\label{1}
\eeq
This Lagrangian is invariant under  $(x,y)\to (-x,-y)$, which implies a double degeneracy of states of energy $E$. It is also  quadratic, with two normal modes: the standard `cyclotronic'  
mode of the $\lambda=0$  model, and a `hyperbolic' mode. The respective angular frequencies are
\beq\label{2}
\omega_c = \frac{eB}{\mu c} \cosh\vartheta,\, \qquad
\omega_h = i\frac{eB}{\mu c} \sinh\vartheta\, , 
\eeq
where $\sinh (2\vartheta) = (2\lambda \mu c^2)/(eB^2)$;  for simplicity, we shall 
consider the limiting case  $\omega_c\gg |\omega_h|$, for which
\beq
\oc \approx  \frac{eB}{\mu c} , \qquad \oh \approx i \frac{\lambda c}{B}\, . 
\label{c8}
\eeq
In the quantum theory, the energy in the cyclotronic mode is 
$(n+ 1/2)\hbar \omega_c$ for integer $n$ labelling the Landau level. 
At sufficiently low energy only the lowest ($n=0$) level is relevant
and the physics is effectively described by the Lagrangian (see e.g.  
\cite{Shankar}) 
\beq
\L_{LLL} =  p\dot x - |\omega_h| xp \, \qquad p= \frac{\hbar y}{\ell^2} \, , \qquad 
\ell^2 = \frac{\hbar c}{eB}\, . 
\eeq
where $\ell$ is the `magnetic' length.  This LLL model is the (unregularized)  one-dimensional $H=xp$ model.  
From the Landau model perspective, each quantum state  is associated  to a quantum of magnetic flux $\Phi_0= 2\pi \hbar c/e$ that occupies an  area $2\pi \ell^2$ in the $xy$-plane.  In the LLL limit,  this  implies the standard  semi-classical quantization  in which one  quantum state is associated with an area $2\pi \hbar$ in phase space. 

The total number of states in each Landau level,  in particular the lowest one, 
is infinite. To  make the number of states in each level finite, 
we shall restrict the particle to the square
\beq\label{allowed}
|x| < L\, , \qquad |y| < L\, . 
\eeq
This implies for the LLL model a restriction on the phase space equivalent to that proposed by Connes for the $H=xp$ model. It also implies that there is a maximum magnitude for the classical energy: $|E|\le L^2/\ell^2$ in units for which $\hbar\omega_h=1$, which  fixes the value of $\lambda$ in terms of $e$ and $B$   by (\ref{c8}). For large $L/\ell$, the semi-classical  estimate for the total  number of quantum states of any energy allowed by this bound is  $L^2/ 2\pi\ell^2$ in each of the four quadrants in the $xy$-plane separated by the $x$ and $y$ axes.  The classical  trajectory for a particle of energy  $E<E_{max}$ is the hyperbola  $xy= E\ell^2$, so the available classical phase space is the region in the square that lies between the two branches of the hyperbola. The number $N_{sc}(E)$ of semi-classical quantum states with energy less than $E$ is the area of this region divided by $2\pi\ell^2$, but because of the double-degeneracy due to the $(x,y)\to (-x-y)$ symmetry we count only those states with $(x>0,y>0)$, which yields
\beq
N_{sc}(E) =  \frac{E}{2 \pi} \log \frac{L^2}{2 \pi \ell^2}
- \frac{E}{2 \pi} \log \frac{E}{2 \pi}  + \frac{E}{2 \pi}\, .
\label{r8}
\eeq
The first term diverges as $L\to\infty$ and may be interpreted as a 
regularization of the continuum of states in the lowest Landau level
for a particle free to move on the entire $xy$-plane. The finite 
correction, which has a magnitude equal to the asymptotic approximation 
of (\ref{asymp}) to the smooth part of the  counting formula, is negative, which
led Connes to interpret it as 
representing spectral lines missing from the continuum.  

\begin{figure}[t!]
\begin{center}
\hbox{\includegraphics[height= 4 cm]{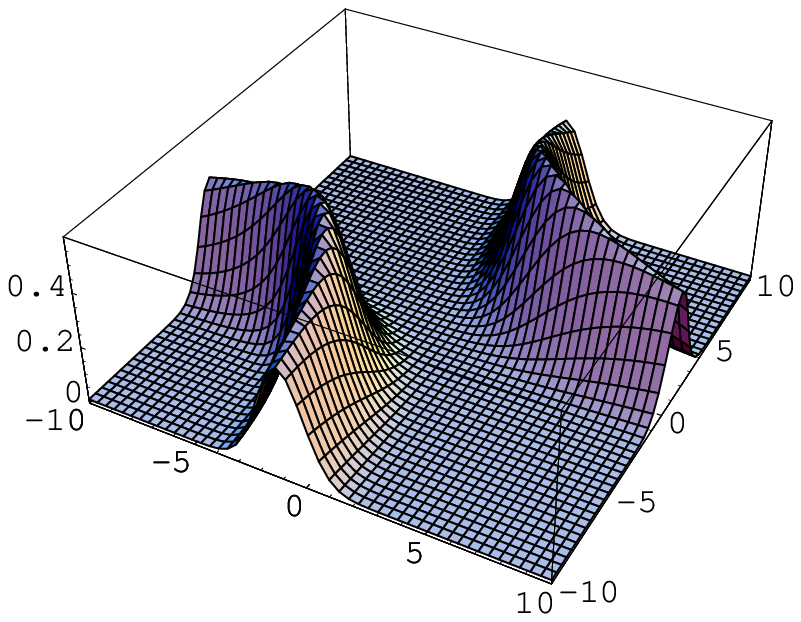}
\includegraphics[height = 4 cm]{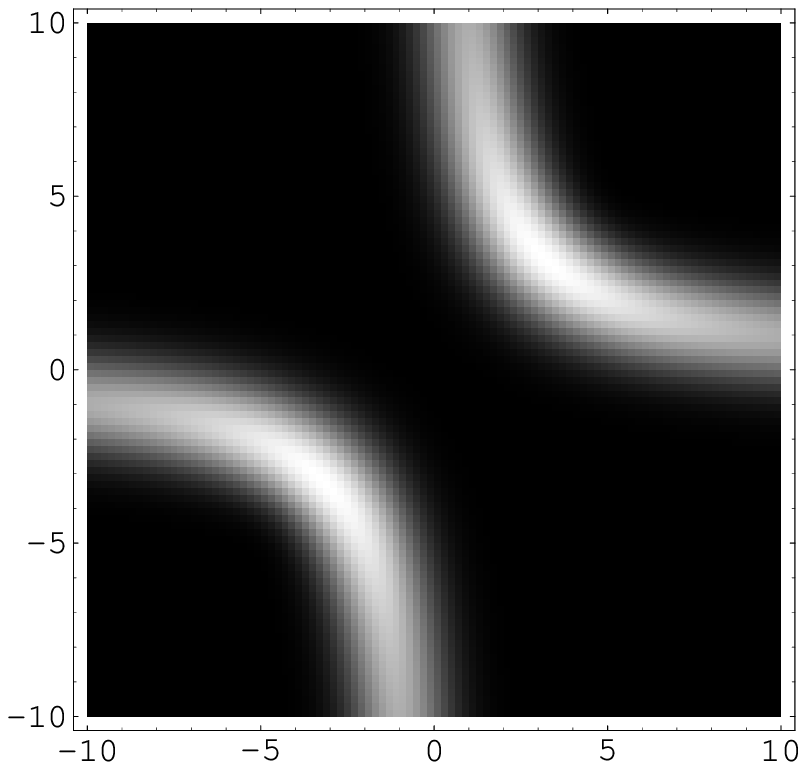}
}
\end{center}
\caption{
Absolute value of the wave function $\psi^+_E(x,y)$
for $E = 10$ in the interval $-10 < x, y < 10$.
The picture on the left gives a 3D representation while that
on the right is a density plot. 
}
\label{psi_3d}
\end{figure}

Another aim of this letter  is to use the Landau model perspective to arrive at a proper
quantum-mechanical understanding of  the regularized $H=xp$ model of Connes; we
focus on that model (details of which may be found in a recent book \cite{CM}) because we have not seen how to achieve the same objective for the Berry-Keating model.    The Hamiltonian operator  corresponding to the starting Lagrangian  (\ref{1}) is
\beq
H = \frac{1}{2 \mu} [\hat p_x^2 + (\hat p_y + \frac{\hbar}{\ell^2} 
x)^2 ] +   e \lambda  x y  \, , 
\label{r51}
\eeq
where $\hat p_x = -i \hbar \partial_x$ and $\hat p_y = -i \hbar \partial_y$. 
There is a unitary transformation in which this Hamiltonian 
becomes the sum of a Hamiltonian $H_c$ for the cyclotronic mode  
and a Hamiltonian $H_h$ for the hyperbolic mode. In units for which 
$\hbar = \ell =1$, one finds that
\beq
H_c = \frac{\omega_c}{2} (\hat p^2 + q^2 ) \, , \qquad
H_h = \frac{|\omega_h|}{2} (Q \hat P + \hat P Q) \, , 
\label{r59}
\eeq
where $\hat P= -id/dQ$ and $\hat p=-i d/dq$. Note the Weyl ordering 
in $H_h$, which follows from the starting Hamiltonian. The unitary
transformation that achieves this decomposition simplifies considerably
in the limit that $\omega_c\gg |\omega_h|$; it corresponds to the classical 
canonical transformation
\beq\label{r60}
q =  x + p_y\, , \quad p = p_x \, , \quad 
Q =  - p_y\, , \quad P = y + p_x \, . 
\eeq
In the limit that $\omega_c\gg|\omega_h|$, any low energy eigenstate 
is the product of the ground state of $H_c$, which is the gaussian
$\psi = \exp (-q^2/2 \ell^2)$, with an eigenstate of $H_h$,  
which we may choose 
to be either even or odd under $Q\to -Q$. These eigenstates of definite
`$Q$-parity'  are
\beq
\Phi_E^+(Q) = \frac{1}{|Q|^{1/2 -iE}}\, , \qquad
\Phi_E^-(Q) = \frac{\sign(Q)}{|Q|^{1/2 -iE}}\, .
\eeq
The corresponding wavefunctions of the original Hamiltonian are 
then given by
\beq
\psi^\pm_E (x,y) =  C \int dQ \,   
 e^{- i Q y/\ell^2} \;e^{- (x -Q)^2/2 \ell^2}\,   \Phi^\pm_E(Q) \, . 
\label{r65}
\eeq
where $C$ is a normalization constant. This yields
\barray
\psi^+_E(x,y) &  = & C^+_Ee^{- \frac{x^2}{2 \ell^2}} \, 
M\left( \frac{1}{4} + \frac{i E}{2}, \frac{1}{2}, 
\frac{ (x - i y)^2}{2 \ell^2} \right) 
\label{l22} \\
\psi^-_E(x,y) &  = & C^-_E (x-iy)e^{- \frac{x^2}{2 \ell^2}} 
\, M\left( \frac{3}{4} + \frac{i E}{2}, \frac{3}{2}, 
\frac{ (x - i y)^2}{2 \ell^2} \right) 
\nonumber 
\earray  
where $M(a,b,z)\equiv {}_1 F_1(a,b,z)$ is a 
confluent hypergeometric  function. A representative 
plot of the even wavefunction is shown in Fig.\ref{psi_3d}. 

\begin{figure}[t!]
\begin{center}
\hbox{\includegraphics[height= 4.1 cm]{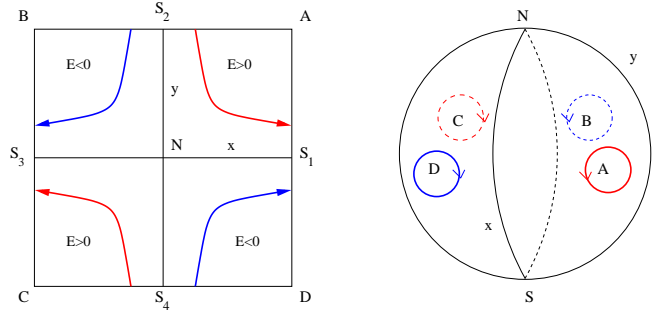}
}
\end{center}
\caption{
Left: $xy$-plane with the classical trajectories 
$x y = E \ell^2$. Right: Sphere that results
from the identifications of the edges. Now the open
trajectories become closed trajectories in one of four segments, enclosing
one of the equatorial points $A,B,C,D$. 
}
\label{xy-plane}
\end{figure}

A classical trajectory that starts  at $(x,L)$ ends  at $(L,x)$, 
so that identification of these points  creates periodic orbits. This identification also means that the square in the  $xy$-plane in which the particle is allowed to move is topologically a sphere, as shown in Fig. \ref{xy-plane}, although classical orbits lie entirely in one of four segments of this sphere. 

For these periodic orbits to emerge from the quantum theory in a 
semi-classical limit, we must identify the wavefunctions at these points, 
up to a phase. To determine the phase we use the asymptotic formulas
\beq
M(a,b,z) \sim \left \{
\begin{array}{ll}
\frac{\Gamma(b)}{\Gamma(a)} \;  e^z \; z^{a-b} & \Re (z) > 0 \\
\frac{\Gamma(b)}{\Gamma(b-a)} \;  (-z)^{-a} & \Re (z) < 0 \\
\end{array}
\right. 
\label{r12}
\eeq
to deduce that 
\begin{eqnarray}
\psi^+_E(L, x)  &\sim&   e^{-ixL/\ell^2- x^2/2 \ell^2}
 \frac{ \Gamma( \frac{1}{2})}{ \Gamma( \frac{1}{4} + \frac{i E}{2}) }
 \; \left( \frac{L^2}{ 2 \ell^2}  \right)^{ - 1/4 + i E/2} \nonumber\\
\psi^+_E(x, L) & \sim &  e^{ - x^2/2 \ell^2} \; 
 \frac{ \Gamma( \frac{1}{2})}{ \Gamma( \frac{1}{4} - \frac{i E}{2}) }
 \; \left( \frac{L^2}{ 2 \ell^2}  
\right)^{ - 1/4 - i E/2} \, .
\label{r14}
\end{eqnarray}
for the even functions (\ref{l22}), and 
\begin{eqnarray}
\psi^-_E(L, x)  &\sim&  L  e^{-ixL/\ell^2- x^2/2 \ell^2}
 \frac{ \Gamma( \frac{3}{2})}{ \Gamma( \frac{3}{4} + \frac{i E}{2}) }
 \; \left( \frac{L^2}{ 2 \ell^2}  \right)^{ - 3/4 + i E/2} \nonumber\\
\psi^-_E(x, L) & \sim & - i \; L    e^{ - x^2/2 \ell^2} \; 
 \frac{ \Gamma( \frac{3}{2})}{ \Gamma( \frac{3}{4} - \frac{i E}{2}) }
 \; \left( \frac{L^2}{ 2 \ell^2}  
\right)^{ - 3/4 - i E/2} \, .
\label{r14bis}
\end{eqnarray}
for the odd functions.   Observe that the `even'  wavefunctions are symmetric, and the `odd' wavefunctions antisymmetric,  under the symmetry operation $(x,y)\to (-x,-y)$.  Counting both symmetric and anti-symmetric functions corresponds, semi-classically, to counting states for $xy>0$, rather than just $(x>0, y>0)$, so we could define the quantum model by considering only the symmetric wavefunctions,  which would effectively mean that we identify $(x,y)$ with $(-x,-y)$. 
The above formulas suggest that we should impose the boundary condition
\beq\label{idenpsi}
\psi^\eta_E(x,L) = 
e^{iLx/\ell^2 + i \pi ( \eta - 1)/4 }  \; \psi^\eta_E(L,x), \;  
\eta = \pm . 
\eeq
Applied to the `even' ($\eta = +$) energy eigenfunctions, 
this leads to the asymptotic condition
\beq
\frac{\Gamma\left(\frac{1}{4} + i\frac{E}{2}\right)}{\Gamma\left(
\frac{1}{4} - i\frac{E}{2}\right)}\, 
\left(\frac{L^2}{2\ell^2}\right)^{-iE} =1\, , 
\eeq
which is equivalent to 
\beq
e^{2i\theta(E)} \left(\frac{L^2}{2\pi\ell^2}\right)^{-iE} =1\, ,
\eeq
where $\theta(E)$ is the function of (\ref{theta}), related to $\bar N(E)$ 
by (\ref{barNTheta}). 
This condition implies that
\beq
\left[\frac{E}{2\pi} \log \left(\frac{L^2}{2\pi \ell^2}\right) + 1 \right] - 
\bar N(E) = N_E
\eeq
where $N_E$ is an integer that we may identify, asymptotically, 
with the number of states with 
energy less than $E$. If the first term on the left hand side is 
interpreted as the regularization of 
the infinite number of states in a continuum, then we see that 
$\bar N(E)$ has the Connes 
interpretation as a count of states missing from this continuum. 
This analysis can be repeated for the odd wave function
$\psi^-_E(x,y)$, in which case the function $\theta(E)$
is replaced by the phase of the odd Dirichlet L-functions.

Finally, we conclude with some speculations on how the fluctuation part of the
Riemann counting formula might arise in the Landau model approach.  
In the context of the 
Berry-Keating model, the  region of phase-space allowed by 
the regularization may be adjusted to `fluctuate'  in such a 
way that  the number of states of positive energy  less than $E$ is 
{\it precisely} $N(E)$  \cite{08-01}. However, the Berry-Keating
model does not seem to arise as the  LLL  limit of a 
model for a particle on a plane,
and the `fluctuating boundary' idea of  \cite{08-01} does not 
work for the Connes model. 
A natural possibility for the Connes model, now viewed as a
lowest Landau limit,  is to suppose that the fluctuation term in the 
Riemann counting formula is related to the higher Landau 
levels. An immediate drawback of this idea is that the full 
Landau model is two-dimensional whereas we need a one-dimensional model. 
We thus seek some  one-dimensional limit of the Landau model that generalizes the standard 
LLL  limit.  To this end, we return to the Hamiltonian  as the sum of $H_c$ and $H_h$, as 
given in (\ref{r59}) and rescale, for convenience, to arrive at the Hamiltonian
\beq
H= \frac{\gamma}{2}\left(\hat p^2 + q^2\right)  + \frac{1}{2}\left( 
Q\hat P + \hat P Q\right)\, , 
\eeq
where $\gamma= \omega_c/|\omega_h|$. Introducing the standard
annihilation operator $a= (1/\sqrt{2})(q+i\hat p)$ and corresponding creation operator, such that
$[a,a^\dagger]=1$, we observe that the operators
\beq
A= aQ^{i\gamma}\, , \qquad A^\dagger = a^\dagger Q^{-i\gamma}
\eeq
have a similar commutation  relation but {\it commute} with the Hamiltonian:
\beq
\left[A,A^\dagger\right] =1 \, , \qquad \left[A,H\right] = 0 = 
\left[A^\dagger,H\right]\, . 
\eeq
Eigenstates of $A$ are coherent states with complex eigenvalues 
$z$. States with $z=0$ are those of the lowest Landau level, 
annihilated by $a$, so we may modify this limit by considering 
eigenstates of $A$ with non-zero eigenvalue. At the classical
level, the equation $A = z$ implies that the cyclotronic motion
is governed  by the hyperbolic one. Taking $z$ real for simplicity,
one finds that
\beq
q = \sqrt{2} z \cos( \gamma \log Q), \, \qquad 
p = - \sqrt{2} z \sin( \gamma \log Q) \; .
\label{qp} 
\eeq
Proceeding semi-classically, we consider the area of phase space enclosed
by a closed classical trajectory, i.e. its action. The total action  receives contributions from 
both modes:
\beq
\CA = \int P \; dQ + \int p \; dq\, . 
\eeq
In the limit, $\gamma >>1$ the first term
on the RHS of this equation gives the estimate (\ref{r8}).
Hence in the same limit and allowing the particle
to gyrate according to (\ref{qp}), one finds the
semiclassical formula 
\begin{eqnarray}
N_{sc}(E) &\sim& \frac{E + \gamma z^2}{2 \pi}  \log \frac{L^2}{2 \pi \ell^2}\nonumber\\
&-&\! \frac{E}{2 \pi} \log \frac{E}{2 \pi}  + \frac{E}{2 \pi} 
- \frac{\gamma z^2 }{2 \pi} \log \frac{E}{2 \pi} \, .
\label{r8bis}
\end{eqnarray}
The first term, which diverges as $L\to\infty$, may be interpreted as a count of states that become a continuum in this limit.  Note the ${\cal O}(\log E)$ correction to the counting of `missing' states, which is exactly the order of  the fluctuation term  in the Riemann counting formula. 

In summary, we have shown that the semi-classical $H=xp$ model,  conjectured to be the semi-classical limit of a quantum mechanical model for the complex zeros of the Riemann zeta function, can be understood as the lowest Landau level limit of a fully quantum mechanical model for a particle on a plane in the presence of electric and magnetic fields. As it stands, a counting of states in the model does not yield the full `counting'  formula for the Riemann zeros because the `fluctuation term' is missing, but we have provided semi-classical evidence that this will arise from a consideration of 
the higher Landau levels. Apart from providing a new tool in the spectral approach to the Riemann hypothesis, there is also the possibility 
that it will allow a laboratory construction of a system for which 
the physics is described by the `Riemann Hamiltonian', the existence of which would prove the Riemann hypothesis.

{\it Acknowledgments-} We  thank M. Asorey, and J. Keating for discussions. 
This work was supported by the  CICYT project FIS2004-04885 (G.S.) and by 
the EPSRC (P.K.T.).  GS also acknowledges ESF Science Programme 
INSTANS 2005-2010.

\end{document}